\documentclass[%
apl,%
 amsmath,amssymb,amsfonts,
reprint,%
]{revtex4-1}

\usepackage{graphicx}% Include figure files
\usepackage{dcolumn}% Align table columns on decimal point
\usepackage{bm}% bold math
\usepackage{epstopdf}
\usepackage{natbib}
\usepackage{float}
\usepackage[mathlines]{lineno}% Enable numbering of text and display math
\usepackage[percent]{overpic} %added by Mehdi
\usepackage{soul}
\begin{document}

\preprint{AIP/123-QED}

\title{Optimum High-k Oxide for the Best Performance \\
of Ultra-scaled Double-Gate MOSFETs }% Force line breaks with \\
%\thanks{Footnote to title of article.}

\author{M. Salmani-Jelodar}
\email{m.salmani@gmail.com.} 
\affiliation{Network for Computational Nanotechnology and School of Electrical and Computer Engineering,
Purdue University, West Lafayette, Indiana, USA 47907}%Lines break automatically or can be forced with \\
\author{H. Ilatikhameneh}%
\altaffiliation{Purdue University}%

\author{S. Kim}%
\altaffiliation{Intel Corp.}%
\author{K. Ng}
\altaffiliation{SRC}%
\author{G. Klimeck}
\altaffiliation{Purdue University}%Lines break automatically or can be forced with \\
 
\date{\today}% It is always \today, today,

\begin{abstract}
A widely used technique to mitigate the gate leakage in the ultra-scaled metal oxide semiconductor field effect transistors (MOSFETs) is the use of high-k dielectrics, which provide the same equivalent oxide thickness (EOT) as $\rm SiO_2$, but thicker physical layers. However, using a thicker physical dielectric for the same EOT has a negative effect on the device performance due to the degradation of 2D electrostatics. In this letter, the effects of high-k oxides on double-gate (DG) MOSFET with the gate length under 20 nm are studied. We find that there is an optimum physical oxide thickness ($\rm T_{OX}$) for each gate stack, including $\rm SiO_2$ interface layer and one high-k material. For the same EOT,  $\rm Al_2O_3$ (k=9) over 3 $\rm\AA$ $\rm SiO_2$ provides the best performance, while for $\rm HfO_2$ (k=20) and $\rm La_2O_3$ (k=30), $\rm SiO_2$ thicknesses should be 5 $\rm\AA$ and 7 $\rm\AA$, respectively. The effects of using high-k oxides and gate stacks on the performance of ultra-scaled MOSFETs are analyzed. While thin oxide thickness increases the gate leakage, the thick oxide layer reduces the gate control on the channel. Therefore, the physical thicknesses of gate stack should be optimized to achieve the best performance. 
%The physics of this phenomena is examined and found the physical oxide thickness by using high-k materials requires to be optimum; Thick enough to keep the gate leakage, low and thin enough to reduce short-channel effects caused by thick oxide layer.  
%
%Valid PACS numbers may be entered using the \verb+\pacs{#1}+ command.
\end{abstract}

\pacs{Valid PACS appear here}% PACS, the Physics and Astronomy
\keywords{High-k dielectric effect, 2D electrostatics, MOSFET Scaling, Double Gate MOSFETs}%Use showkeys class option if keyword
\maketitle

\textbf{INTRODUCTION} - The scaling of transistors requires the thinning of $\rm SiO_2$ gate oxide \cite{ITRS}, which can induce significant gate tunneling below 1 nm oxide thicknesses. To mitigate the gate tunneling current in a thin oxide layer of ultra-scaled MOSFETs, high-k dielectrics are used \cite{IntelFinFET30nm, Salmani2014APL}. However, due to the thicker high-k gate oxides, performance drops have been observed in the ultra-scaled MOSFETs with k $>$ 30 \cite{xie2012review, xie2013comprehensive, Salmani2014SNW, cheng1999impact, frank1998generalized, manoj2007impact}. 
Thicker $\rm T_{OX}$ from larger k values worsen the short channel effects, even if the EOT is kept the same \cite{xie2012review}. This happens due to the effect of the lateral field in the oxide, which is more pronounced in higher k materials \cite{xie2012review, xie2013comprehensive}, and fringing capacitance due to spread of potential between the gate and the source and drain \cite{Salmani2014SNW}.
It is known that not only EOT, but also the physical oxide thickness plays important roles in SCEs. However, it is not clear what the solution is as the gate lengths of MOSFETs approach below 20 nm \cite{ITRS, Salmani2014APL, Samsung2011Bulk}. 

In this work, we attempt to answer these questions: 
1) What is the impact of using a different high-k materials with the same EOT on an ultra-scaled DG MOSFET? 
2) What is the optimum thickness of high-k gate stack for a fixed EOT?  
3) How do we analytically estimate the gate leakage in the off state for a specific gate stack?
We show that for ultra-scaled DG MOSFETs, there is an optimum oxide thickness that balances gate leakage of a thin oxide layer and SCEs of thick oxides. We provide an analytical model for the effect of high-k on the gate control and gate leakage current.

\textbf{METHODOLOGY} - A self-consistent Schr\"{o}dinger-Poisson solver, based on the real-space effective-mass approximation and the wave function formalism, including direct gate and Fowler-Nordheim tunneling currents is used to simulate DG MOSFETs \cite{luisier2007three, luisier2008two,steiger2011nemo5, fonseca2013efficient}. This work only considers electrons for calculations, while holes are not taken into account for the n channel devices. Neglecting holes means that band-to-band tunneling is ignored in these simulations. This effect is negligible in devices, whose source-to-drain voltage ($\rm V_{DS}$) is smaller than the band gap of the channel material \cite{luisier2006atomistic, salmaniTunneling}.
DG device specifications are in correspondence to the ITRS table data for the 2015 node (Fig. 1). The transport direction is aligned with the $<$110$>$ crystal axis and the confinement direction is (001). The source and drain regions are doped with a donor concentration ND =  $\rm 10^{20} [/cm^3$]. For all simulations EOT is fixed at 0.86 nm. The gate dielectric constant varies from 3.9 to 30 (Table I). Physical oxide thickness ($T_{OX}$) changes with the k value to keep EOT fixed, according to the following equation for the cases without any interface layer:
\begin{eqnarray}
T_{OX} = EOT \times \frac{k_{OX}}{k_{SiO_2}}
\end{eqnarray}
However, for the gate stacks with $\rm SiO_2$ interface layer and fixed EOT, $T_{OX}$ = $\rm T_{SiO_2}$+$\rm T_{high-k}$ and $\rm T_{high-k}$ is calculated as:
\begin{eqnarray}
T_{high-k} = (EOT-T_{SiO_2}) \times \frac{k_{high-k}}{k_{SiO_2}}
\end{eqnarray}
Two gate-dielectric configurations with five different oxide materials are examined in this paper. (I) All oxide materials are directly deposited on the Si channel, without any interface oxide, with the same EOT, but different k and $T_{OX}$. TiN metal gate contacts are characterized by work function of 4.25 eV, and their electron effective mass ($m^* = m_0$). (II) $\rm SiO_2$-high-k gate stack with TiN contacts. Effective masses are assumed isotropic (Table I). Relative dielectric constants and band gap for each material are described in Table I. Sub-threshold swing (SS), drain induced barrier lowering (DIBL) and gate tunneling current are calculated at the OFF-state, where drain off current is kept at about 100 nA/um. %ON-current is the drain current of the device with $\rm V_{DS} = V_{DD}$ and $\rm V{GS} = V_{GS_{@OFF}}+V_{DD}$, where $\rm V_{GS_{@OFF}}$ is the gate voltage at drain current equal to 100 nA/um.

\begin{figure}[h]%{1}
\begin{center}
%\begin{tabular}[c]{cc}
%	\begin{tabular}[c]{c}
		\begin{overpic}[width=\columnwidth]{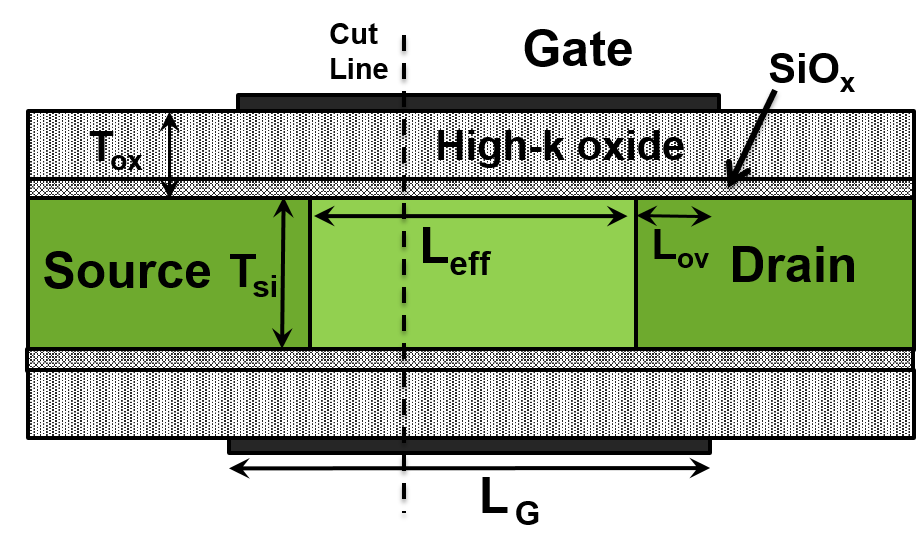} %FINAL_Lg_Scaling_V2012.jpg}% [width=0.47\textwidth][width=10cm,grid,tics=10] width=0.48\textwidth Scaling_Lg_VDD_EOT.jpg
%		\put(25,73.5){(a)}
		\end{overpic}
%	\end{tabular} & \\ 
%\end{tabular}
\end{center}
\vspace{-2em}
\caption{\label{fig:Geometry} 
Schematic view of the double-gate MOSFETs considered in this paper. The geometry parameters are the same node 2015 from ITRS tables \cite{ITRS}. The gate length is set to 16.7 nm. There is gate-source/drain overlap equal to 10$\%$ of the gate length (i.e. 1.6 nm). The thickness of Si channel is 5.3 nm. Currents are normalized by the channel width per side, which means simulated current divides by 2. Supply voltage ($\rm V_{DD}$) is set to 0.83 V.
}
\vspace{-2em}
\end{figure}

\begin{table}[h]
  \begin{center}
  	\caption{Dielectric constant k, band gap, conduction band (CB) offset to Si, and tunneling effective mass ($m^*$) for the oxide materials used in this work are shown. tunneling effective masses are from multiple experiments and theoretically calculated references \cite{yeo2002direct, robertson2000band}. P values are calculated from equation 4 for each specific oxide material. Starred (*) numbers are used for these materials in our simulations. Physical thickness of each oxide material is shown for both configurations I and II.  }
    \begin{tabular}{| l | c c c c c |}
    \hline
	Material & $SiO_2$ & $Si_3N_4$	& $Al_2O_3$ &	$HfO_2$	& $La_2O_3$  \\ \hline   
K & 3.9	&7*-8	& 9*-10	& 20*-25 &	27-30* \\ \hline
    Band gap [eV]&	9	& 5.3	& 8.8	&5.8	&6 \\ \hline
    CB offset [eV]&	3.5	& 2.4	& 2.8	& 1.5	& 2.3 \\ \hline
    $m^* /m_0$	&0.4*-0.5	&0.4	&0.35	&0.11*-0.17	&0.27 \\ \hline
    P[$nm^{-1}$] & 6.06	& 5.02	& 5.07	& 2.08 & 4.04  \\ \hline
    $\rm T_{OX} (No \: Interface) $ &	0.9	& 1.6	& 2.0	&  4.9	& 6.6\\ \hline 
    $\rm T_{OX} (T_{SiO_2} = 3 \AA) $ &	-	& 1.3	& 1.6	&  3.5	& 4.6\\ \hline
    $\rm T_{OX} (T_{SiO_2} = 5 \AA) $ &	-	& -	& 1.4	&  2.5	& 3.3\\ \hline
    $\rm T_{OX} (T_{SiO_2} = 7 \AA) $ &	-	& -	& 1.1	&  1.5	& 1.9\\ \hline    
    \end{tabular}
  \end{center}
\vspace{-1.5em}
\end{table}

\textbf{RESULTS AND DISCUSSION} – The ballistic $\rm I_D-V_{GS}$ characteristics of DG MOSFET presented in Fig. 1 are simulated for $\rm V_{DS}$ = $\rm V_{DD}$ = 0.83 V. 
Fig. 2-A shows the results for the configuration I, without interface layer and with no gate leakage assumption. As expected, the thicker oxide would degrade the device performance by increasing the sub-threshold swing (SS) and lowering ON-current for the fixed OFF-current \cite{Salmani2014SNW}. However, when gate tunneling is included, in the $\rm SiO_2$ case, the OFF-current rises above 100 nA/um. OFF-current below 100 nA/um is considered as the OFF state in ITRS guideline \citep{ITRS}. For $\rm Si_3N_4$, in the configuration I, ON-current is around 3430 $\rm uA/um$, which is a bit less than ON-current for the device with $\rm Al_2O_3$ oxide. Although, $\rm T_{OX_{Si_3N_4}}$ is thinner than $\rm T_{OX_{Al_2O_3}}$, it provides less ON-current for the fixed OFF-current. This is a result of high gate leakage at OFF state in $\rm Si_3N_4$, which drops SS as well. $\rm I_{Gate_{Si_3N_4}}$ is 90 times more than  $\rm I_{Gate_{Al_2O_3}}$ and contributes as $ 22\: \%$ of $\rm I_{OFF}$. As the gate leakage cannot be controlled by gate voltage, it degrades sub-threshold swing, and consequently the ON-current.

Figs. 2-B to 2-D show the effect of using interface layer and its thickness on $\rm I_D-V_{GS}$. Fig. 2-B depicts the transfer characteristics of a device with $\rm Al_2O_3$ as the gate stack. As it is shown in Table I, $\rm T_{OX}$ ($\rm T_{OX}$ = $\rm T_{SiO_2}$ + $\rm T_{Al_2O_3}$) with different interface layer thickness up to 5 $\rm \AA$, is still thick enough to keep the gate leakage very low. Thin high-k gate stack (4 $\rm \AA \: Al_2O_3$ on top of 7 $\rm \AA$ $\rm SiO_2$) cannot keep the leakage current low enough to turn off the device. The electrostatics and the gate leakage in the required voltage range does not vary drastically, which keep the $I_D-V_{GS}$ characteristics similar in the interested range. In Fig. 2-C high-k material is $\rm HfO_2$. Reduction in the $\rm T_{OX}$ from 4.9 nm (no interface layer) to 1.5 nm (7 $\rm \AA$ interface layer) impacts on SCEs. All cases can turn off the device ($\rm I_{OFF} <$ 100 nA/um) except the case with 7 $\rm \AA$ interface layer. The case with 5 $\rm \AA$ interface layer provides the best SS and better ON current at fixed $\rm I_{OFF}$ of 100 nA/um. In the case of $\rm La_2O_3$ as high-k material in Fig. 2-D, $T_{OX}$ reduces from 6.6 nm (no interface layer) to 1.9 nm (7 $\rm \AA$ interface layer) impacts SCEs drastically. Each case turns off the device properly ($\rm I_{OFF} <$  100 nA/um), however the case with 7 $\rm \AA$ interface layer provides the best $\rm I_{ON}$ at fixed $\rm I_{OFF}$, as it has the thinnest oxide. 

\begin{figure}[h]%{1}
\begin{center}
\begin{tabular}[c]{cc}
   \begin{tabular}[t]{c}
   %\centering header t
		\begin{overpic}[width=\columnwidth]{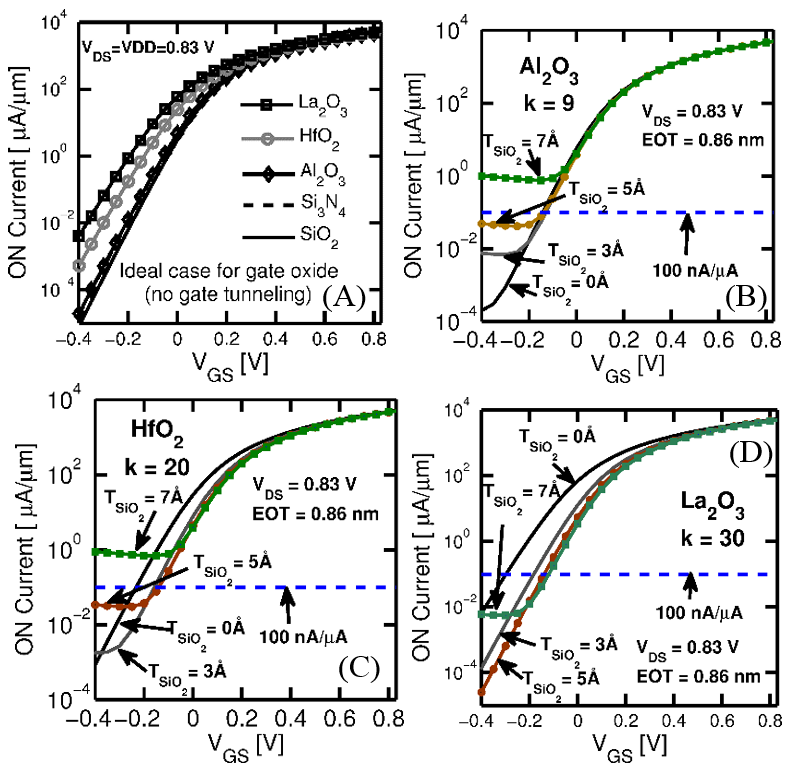}
		%\put(10,75){(a)} 
		\end{overpic}	
  \end{tabular} & \\ 
  
\end{tabular}
\end{center}
\vspace{-2em}

\caption{\label{fig:IV} 
Typical $I_D-V_{GS}$ transfer characteristics at $\rm V_{DS}$ = 0.83 V of DG MOSFETs with fixed EOT of 0.86 nm, and various dielectric materials (different k). (A) With the assumption that ideal oxides which do not have any leakage. (B) with $\rm Al_2O_3$ and $\rm SiO_2$. (C) with $\rm HfO_2$ and $\rm SiO_2$. (D) with $\rm La_2O_3$ and $\rm SiO_2$.}
\end{figure}

In Figs. 3-A to 3-D, performance metrics, including $I_{ON}$, gate leakage current in the OFF-state, subthreshold swing (SS) and DIBL for fixed $\rm I_{OFF}$ of 100 nA/um are depicted. $I_{ON}$ increases by reduction of the high-k thickness for the same EOT in the gate stack (i.e. increase in $\rm SiO_2$ interface layer thickness). Optimum physical oxide thickness helps the device to turn off, as well as provides stronger electrostatics. The optimum oxide thickness for each gate stack can be achieved with a layer of $\rm SiO_2$ and high-k material. $\rm Al_2O_3$ gate stack shows optimum ON current at $\rm T_{ SiO_2}$ = 3 $\rm \AA$. However, for both $\rm HfO_2$ and $\rm La_2O_3$ oxide materials, thicker $\rm SiO_2$ interface layer improves the device $I_{ON}$ (Fig. 3-A). In Fig. 3-B, the gate leakage current at the OFF-state is shown, which is part of the drain current. The drain current has two components; the source to drain current and the gate to drain current (gate leakage). Gate leakage results from the tunneling of electrons through the potential barrier between the gate and the channel. $\rm I_{Gate}$ is exponentially related to the oxide thickness ($T_{OX}$ = $\rm T_{SiO_2}$+$\rm T_{high-k}$) and oxide effective mass \citep{Salmani2014APL, salmaniTunneling}. $\rm Al_2O_3$ and $\rm HfO_2$ have similar $I_{Gate}$. This similarity in $I_{Gate}$ is a result of the light effective mass of $\rm HfO_2$ (0.11 $\rm m_0$) compared to $\rm Al_2O_3$ (0.35 $\rm m_0$), while their  dielectric constants are very different. Replacing oxide materials and varying the interface layer thickness impacts SS values (Fig. 3-C). SS depends on the current at sub-threshold region, which is controlled by the electrostatics and the gate leakage \cite{salmaniTunneling}. Thinning $T_{OX}$ enhances the gate control, which improves SS until $\rm I_{Gate}$ becomes comparable to $\rm I_{OFF}$. $\rm I_{Gate}$ does not change exponentially with the gate bias, which leads to higher SS value. As it is depicted in Fig. 3-D DIBL improves by a reduction in the oxide thickness \cite{Salmani2014SNW}, but for very thin oxides, higher gate leakage slows down the DIBL value reduction [9]. Thin $\rm T_{OX}$ improves the gate control over the channel (or top of the barrier), which results in weaker drain control over the channel. 

\begin{figure}[h]%{1}
\begin{center}
\begin{tabular}[c]{cc}
  \begin{tabular}[c]{c}
		\begin{overpic}[width=\columnwidth]{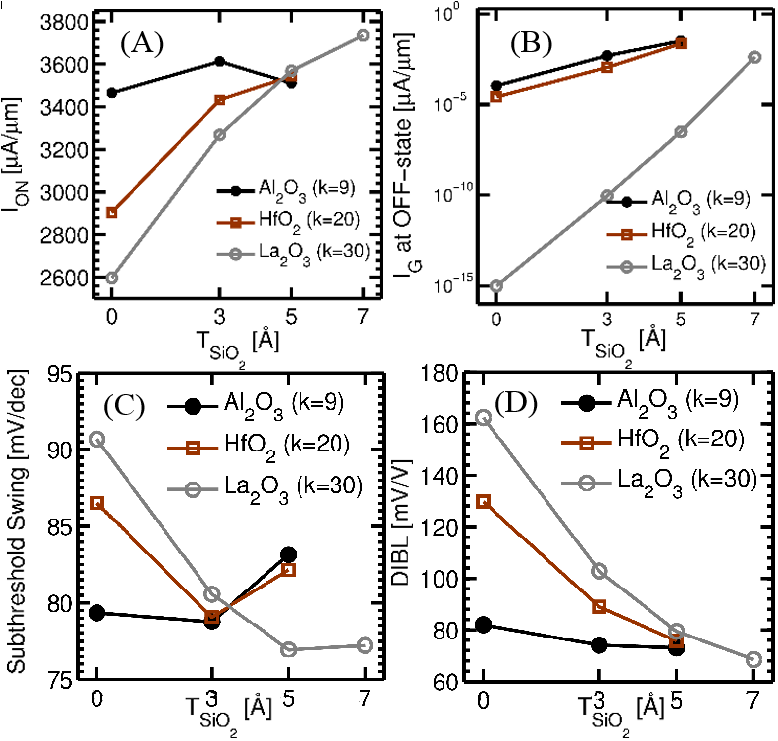}%FINAL_TunnelingDGHP2013_LG.jpg} %TunnelingDGHP2015.jpg [scale=0.8]
%		\put(22,89){(a)} 
		\end{overpic}

  \end{tabular} 
\end{tabular}
\end{center}
\vspace{-2em}
\caption{\label{fig:RESULTS} 
Performance metrics of DG MOSFETs with fixed EOT of 0.86 nm, and various dielectric materials (different k) at fixed $I_{OFF}$ (100 nA/um). If $\rm I_D-V_{GS}$ could not go below 100 nA/um at off state were dropped out of these figures. (A) ON current $I_{ON}$, (B) Gate leakage ($I_{Gate}$), (C) Sub-threshold swing (SS) and (D) Drain induced barrier lowering (DIBL).}
\end{figure}

Gate leakage occurs as the carrier tunnels through the oxide layer between the gate contact and the channel. The potential barrier for a gate stack is depicted in Fig. 4-A. Using the tunneling transmission equation, we can estimate the tunneling gate leakage. In Fig. 4-B, the tunneling transmission is calculated using equation 3, which shows a strong correlation with gate leakage in Fig. 2-B. Tunneling transmission (Trans) for the gate stack is calculated as:
\begin{eqnarray}
Trans \approx e^{-2(T_1.P_1+T_2.P_2+...)}
\end{eqnarray} 
\begin{eqnarray}
P = \sqrt{\frac{2m^*U}{\hbar^2}}
\end{eqnarray} 
where $T_i$ and $P_i$ are thickness and decaying wave-vector of carrier in $i^{th}$ oxide within the gate stack, accordingly. The decaying wave-vector, $P_i$, in each oxide layer is calculated from the effective mass, $m^*$, and the potential barrier height, $U$ (the CB offset in Table I), which are listed in Table I for different high-k materials. Current is calculated as $I \approx  \frac{q^2}{h} .  M . Trans . dU$, where $M$ is number of modes, q is electron charge and h is Planc constant and dU is tunneling energy window, which is equal to the potential difference between channel and gate. Except $Trans$, which is calculated from equation 3 the rest of the relation for current is fixed for all of the DG devices in this work \cite{datta2005quantum}. %\frac{q^2}{\hbar}
\begin{figure}[h]
\begin{center}
\begin{tabular}[c]{c}
  \begin{tabular}[t]{c}
		\begin{overpic}[width=0.47\columnwidth]{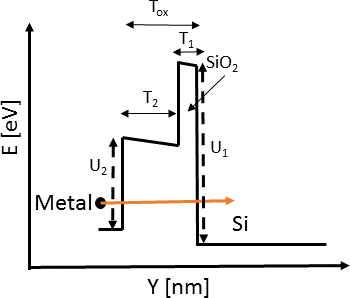}
		\put(12,68){(A)} 
		\end{overpic}
		 \begin{overpic}[width=0.47\columnwidth]{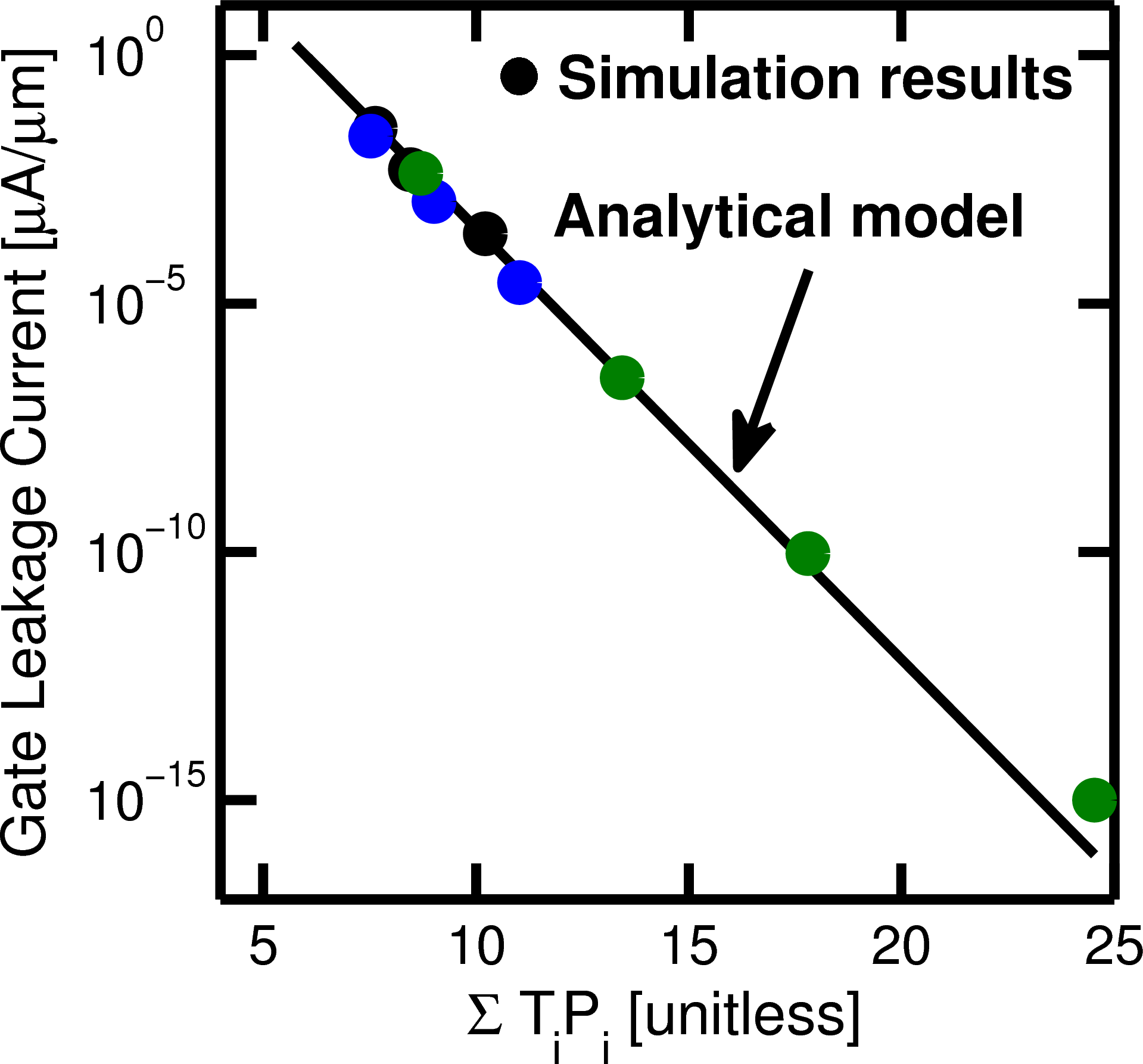}
		\put(22,68){(B)} 
		\end{overpic} \\           
  \end{tabular} 
\end{tabular}
\end{center}
\vspace{-2em}
\caption{\label{fig:tunnelingANALYSIS}
(A) tunneling through the potential barrier for a 2-material gate stack. (B) Gate tunneling current calculated by the presented extensive computational model, QTBM, and the line is calculated by the analytical model from equation 3 and $I \approx  \frac{q^2}{h} .  M . Trans . dU$. Different colored circles are the results from rigorous simulations for different gate stacks.}
\vspace{-1em}
\end{figure}
 
In Figs. 5-A to 5-D, potential difference of OFF and ON states are shown overlapped with the electron flow in the OFF state for $\rm SiO_2$ (Fig. 5-A) and different $\rm HfO_2$ thicknesses over the interface layers. The potential difference between ON and OFF states shows only the effect of gate voltage variation on the potential and removes the potential difference between source and drain. %Potential difference over source/drain results in fringing fields ($E = dV/dr$).  THIS SENTENCE IS WRONG 
Potential spread over the source and the drain are larger for thicker oxides (Figs. 5-A and 5-B). These potential spreads show the effect of increasing fringing capacitance by using thicker oxides, which degrades the device performance. By $\rm T_{OX}$ reduction, the potential spread reduces (Figs. 5-B to 5-E). However, in the very thin oxide case (Fig. 5-E) gate tunneling increases, which drastically reduces the device performance. 

\begin{figure}
\begin{center}
\begin{tabular}[c]{c}
  \begin{tabular}[t]{c}
		\begin{overpic}[width=0.9\columnwidth]{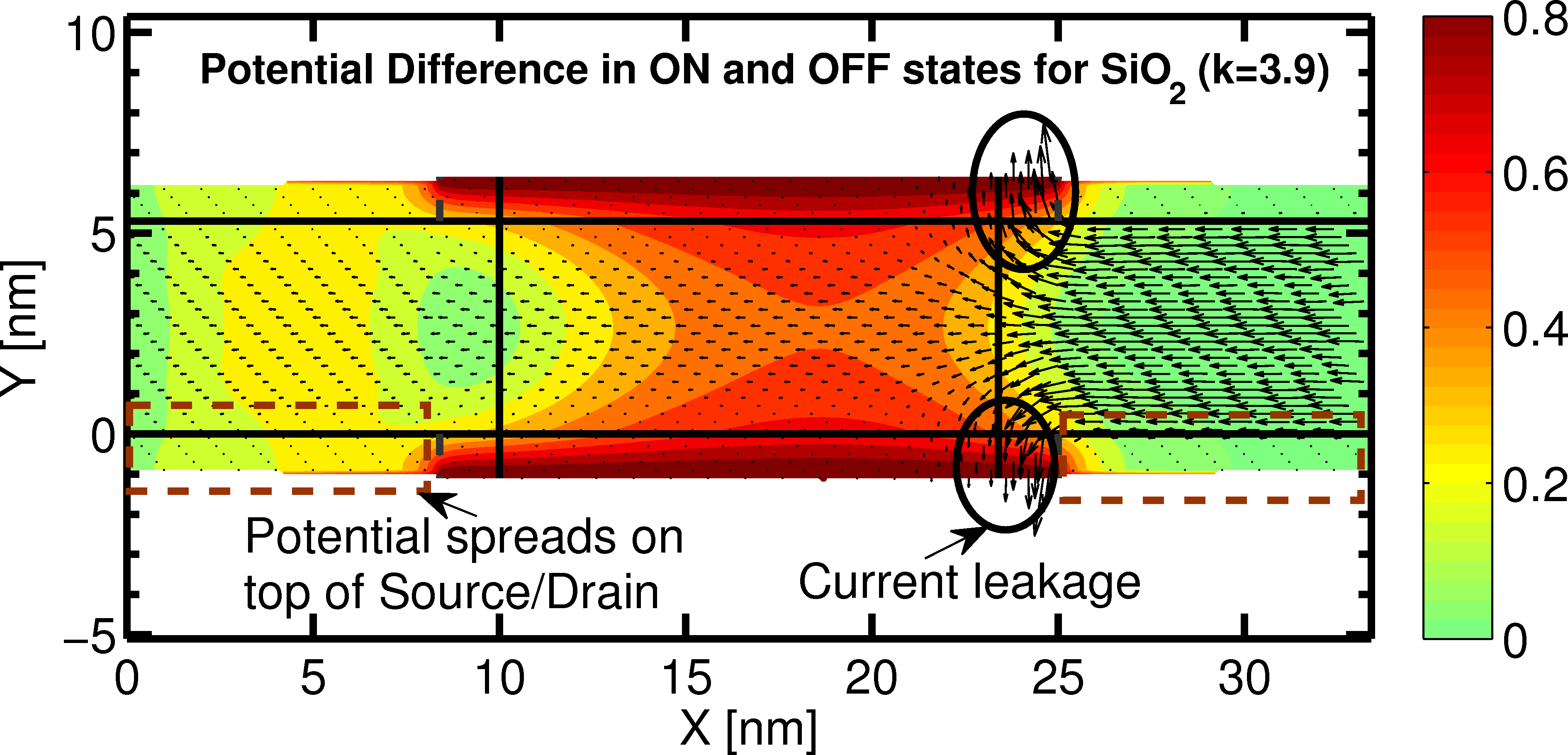} %0.9
		\put(-2,38){(A)} 
		\end{overpic}\\
		 \begin{overpic}[width=0.9\columnwidth]{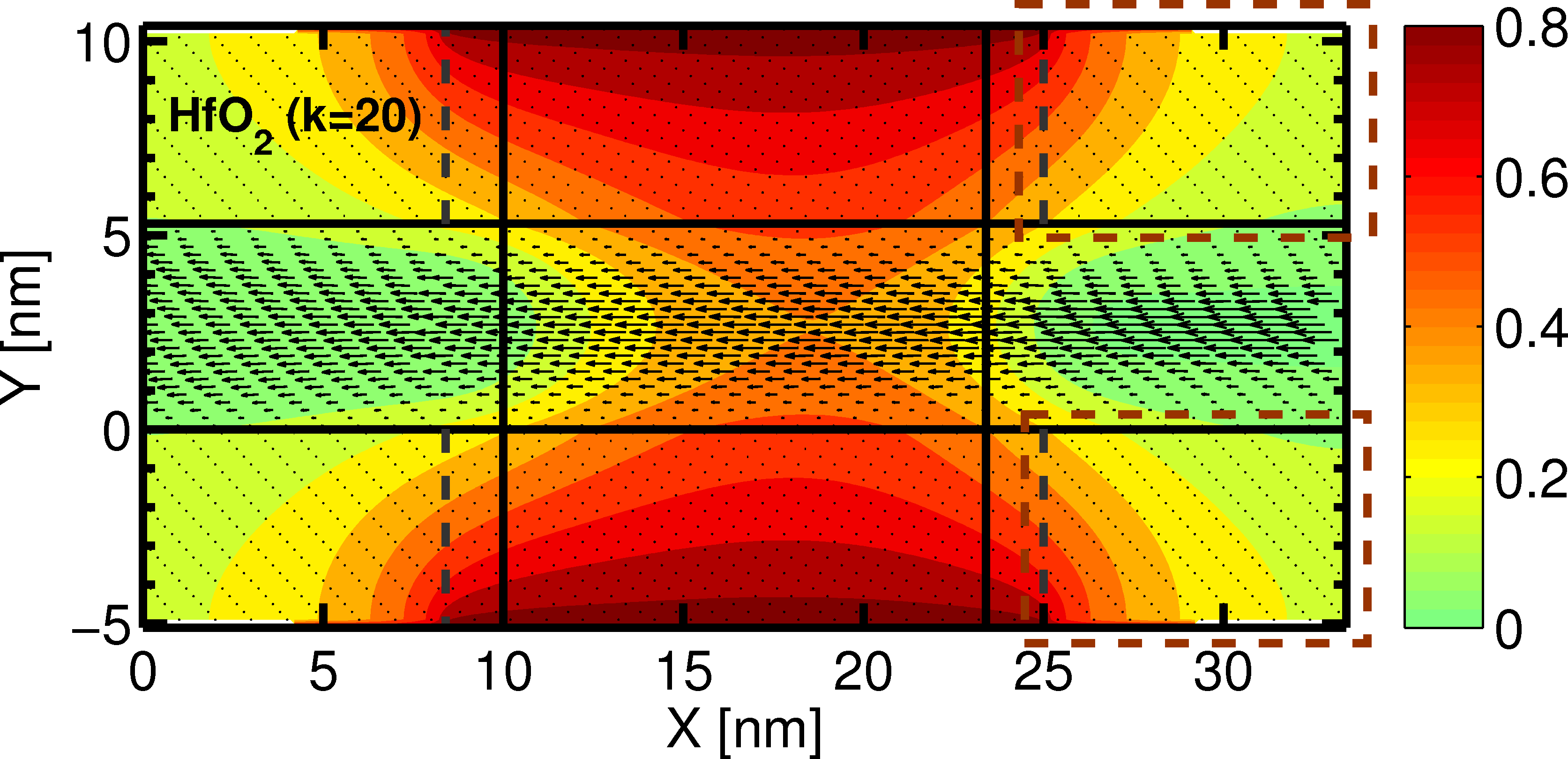}
		\put(-2,38){(B)} 
		\end{overpic} \\           
		 \begin{overpic}[width=0.9\columnwidth]{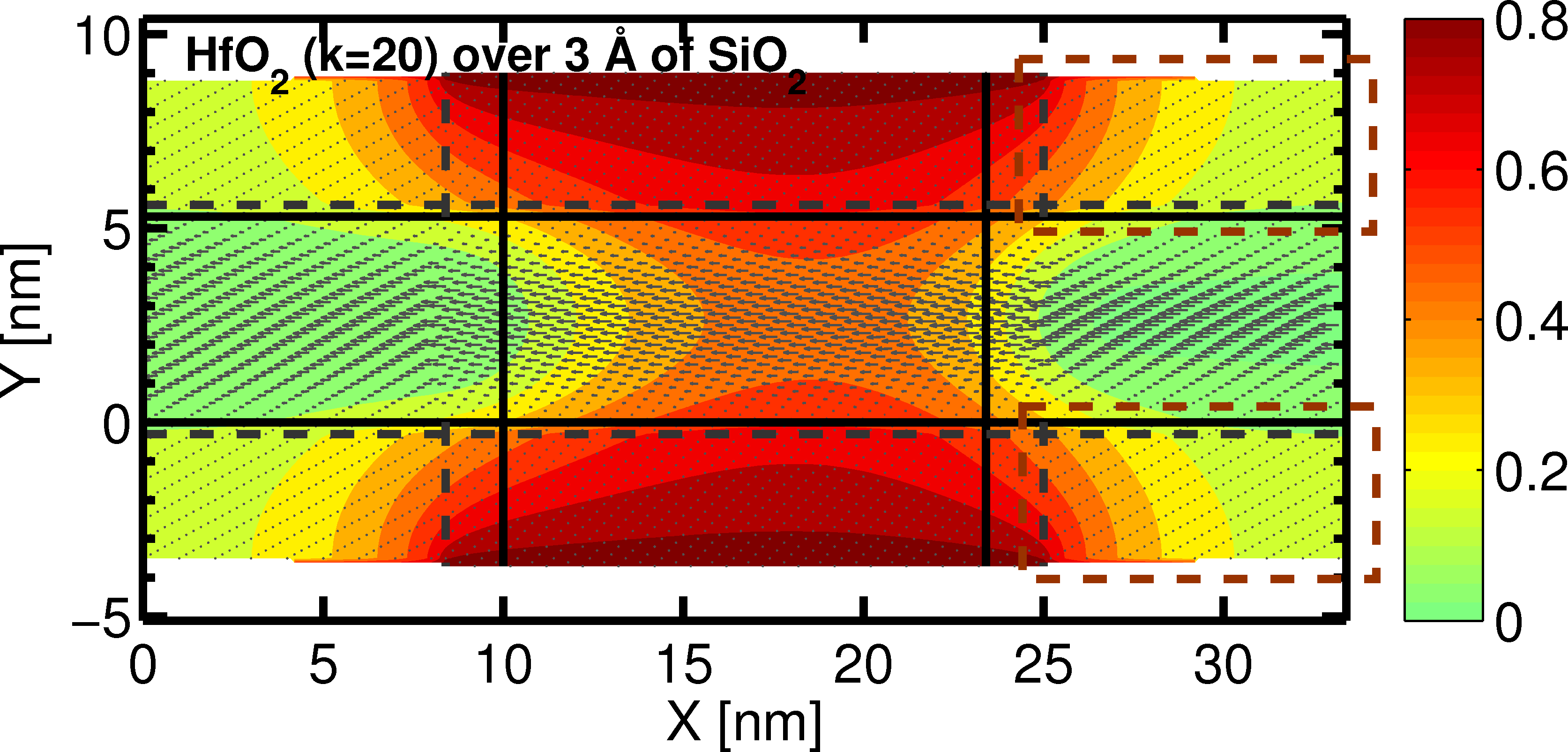}
		\put(-2,38){(C)} 
		\end{overpic} \\           
		 \begin{overpic}[width=0.9\columnwidth]{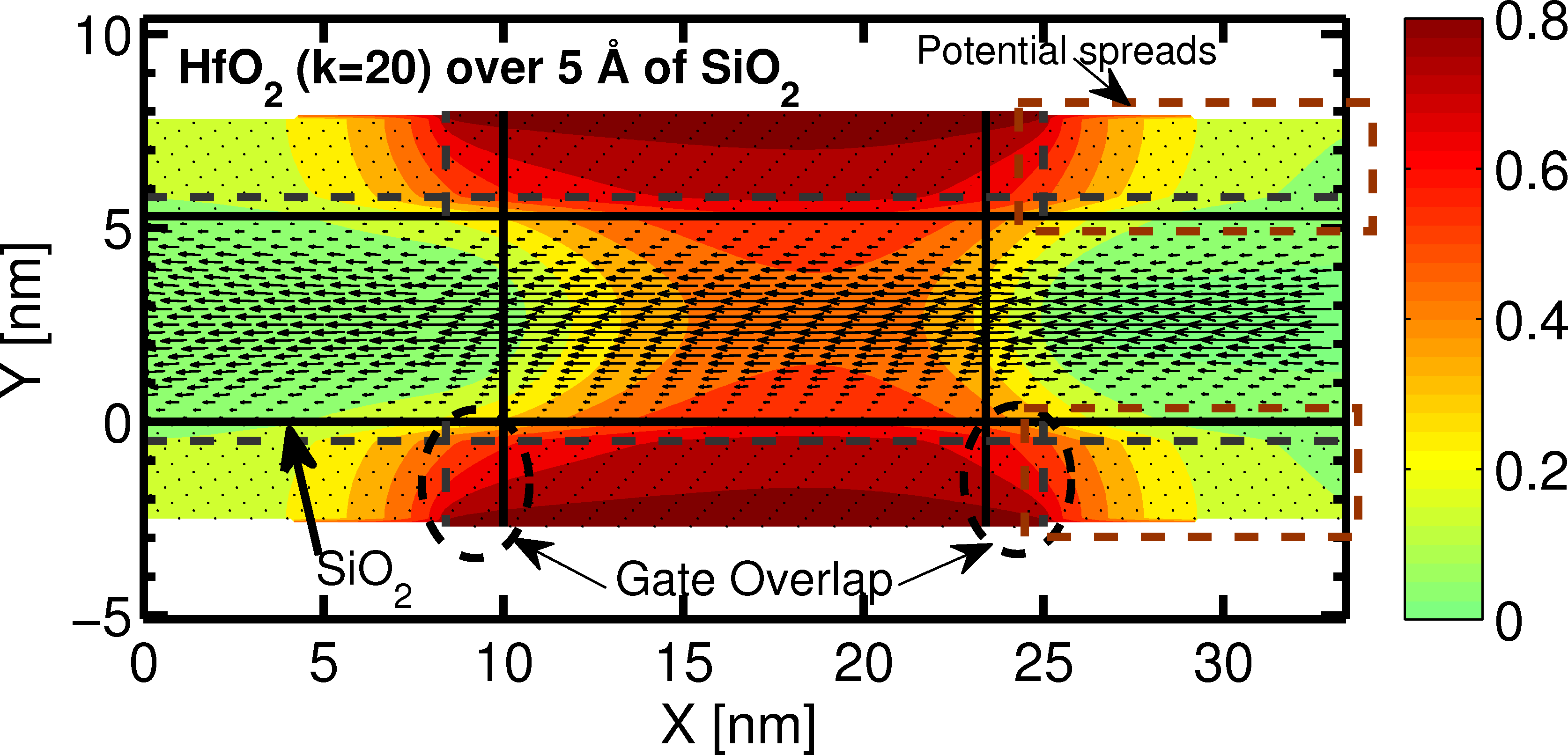}
		\put(-2,38){(D)} 
		\end{overpic} \\           
		 \begin{overpic}[width=0.9\columnwidth]{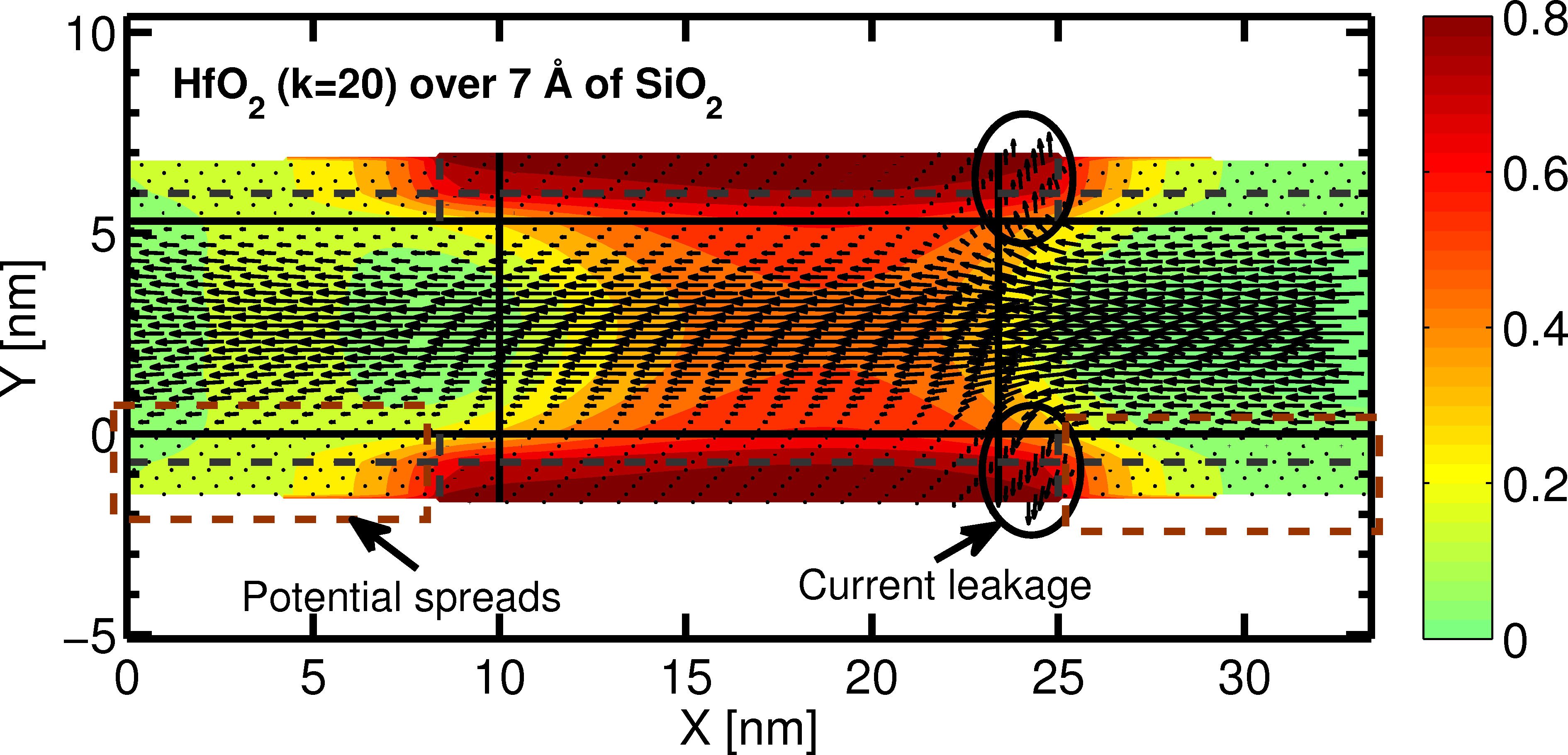}
		\put(-2,38){(E)} 
		\end{overpic}  \\          
        %% Put figure (b) here
  \end{tabular} 
\end{tabular}
\end{center}
%\vspace{-2em}
\caption{\label{fig:POTPROFILE} 
The potential difference between OFF and ON states overlapped with the electron flow in the OFF-state for DG MOSFET with (A) $\rm SiO_2$ oxide. (B) $\rm HfO_2$ without interface layer. (C) $\rm HfO_2$ over 3 $\rm \AA$ interface layer. Potential spread over the source and the drain sections are weaker than only $\rm HfO_2$ case, but stronger than $\rm SiO_2$.  (D) $\rm HfO_2$ over 5 $\rm \AA$ interface layer. Potential spread over the source and the drain sections are weaker than $\rm HfO_2$ over 3 $\rm \AA$ interface layer, but stronger than $\rm SiO_2$. Still, gate tunneling is negligible. (E) $\rm HfO_2$ over 7 $\rm \AA$ interface layer. Gate leakage current is very strong which prevents the device from turning off.}
\end{figure}

Fig. 6-A shows the capacitance network, including the fringing capacitance between the gate and the source/drain, which will be added to the gate capacitance. From the potential profile and fringing field in the Figs 5-A to 5-D, it can be observed that the potential profile broadening is proportional to the physical oxide thickness. This degrades oxide capacitance ($\rm C_{OX}$). We found that the total oxide capacitance can be crudely estimated as:
\begin{eqnarray} 
 \frac{C_{OX}}{A} \approx \frac{\epsilon_{OX}}{T_{OX}}(1-\frac{\alpha T_{OX}}{L_G})
\end{eqnarray}  
where A and $\rm L_G$ are the gate area and the gate length, and $\epsilon_{OX}$ and $\alpha$ are the equivalent gate stack dielectric constant and the empirical factor to estimate $\rm C_{OX}$ degradation. In double gate, it is found that $\alpha$ to be around 0.5. Fig. 6-B shows the effect of the fringing capacitance on the total gate oxide capacitance from our simple model $\frac{\epsilon_{OX}}{T_{OX}}(1-\frac{\alpha T_{OX}}{L_G})$ on the SS values. When $\rm SiO_2$ interface layer is thin, the physical thickness and fringing capacitance are large and the fringing capacitance deteriorates the performance for thick gate stacks. This simple model can be used to \textit{estimate} the effect of physical oxide thickness and gate length on the performance of DG MOSFETs. Here, we used our analytical model to estimate SS using the following equation \cite{sze2006physics}:
\begin{eqnarray}
SS \approx ln(10)\frac{kT}{q}(1+\frac{C_{Si}}{C_{OX}})
\end{eqnarray}  
where $C_{Si}$ is the channel capacitance and $C_{OX}$ is the modified oxide capacitance from equation 5 that takes into account fringing field effect. In Fig. 6-B, $C_{Si}$ is extracted from the calculated SS for the DG MOSFET with only $\rm La_2O_3$ as oxide. Then, SS for gate stack with different $\rm SiO_2$ interface layer is estimated using equations 5 and 6. As $\rm T_{SiO_2}$ increases and $\rm T_{OX}$ decreases, the gate tunneling begins to deteriorate the performance and saturates SS, which shows there are other factors such as gate leakage, and source to drain tunneling \cite{Salmani2014APL} in determination of SS and equation 6 will not work well at that point.

\begin{figure}
\begin{center}
\begin{tabular}[c]{c}
  \begin{tabular}[t]{c}
		\begin{overpic}[width=0.54\columnwidth]{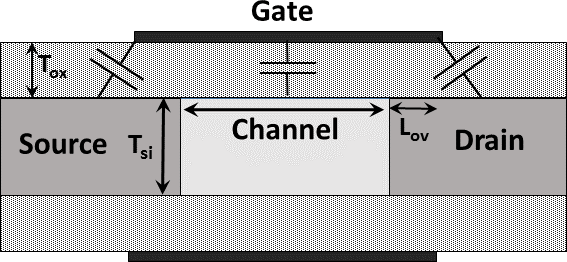}
		\put(2,48){(A)} 
		\end{overpic}
		 \begin{overpic}[width=0.4\columnwidth]{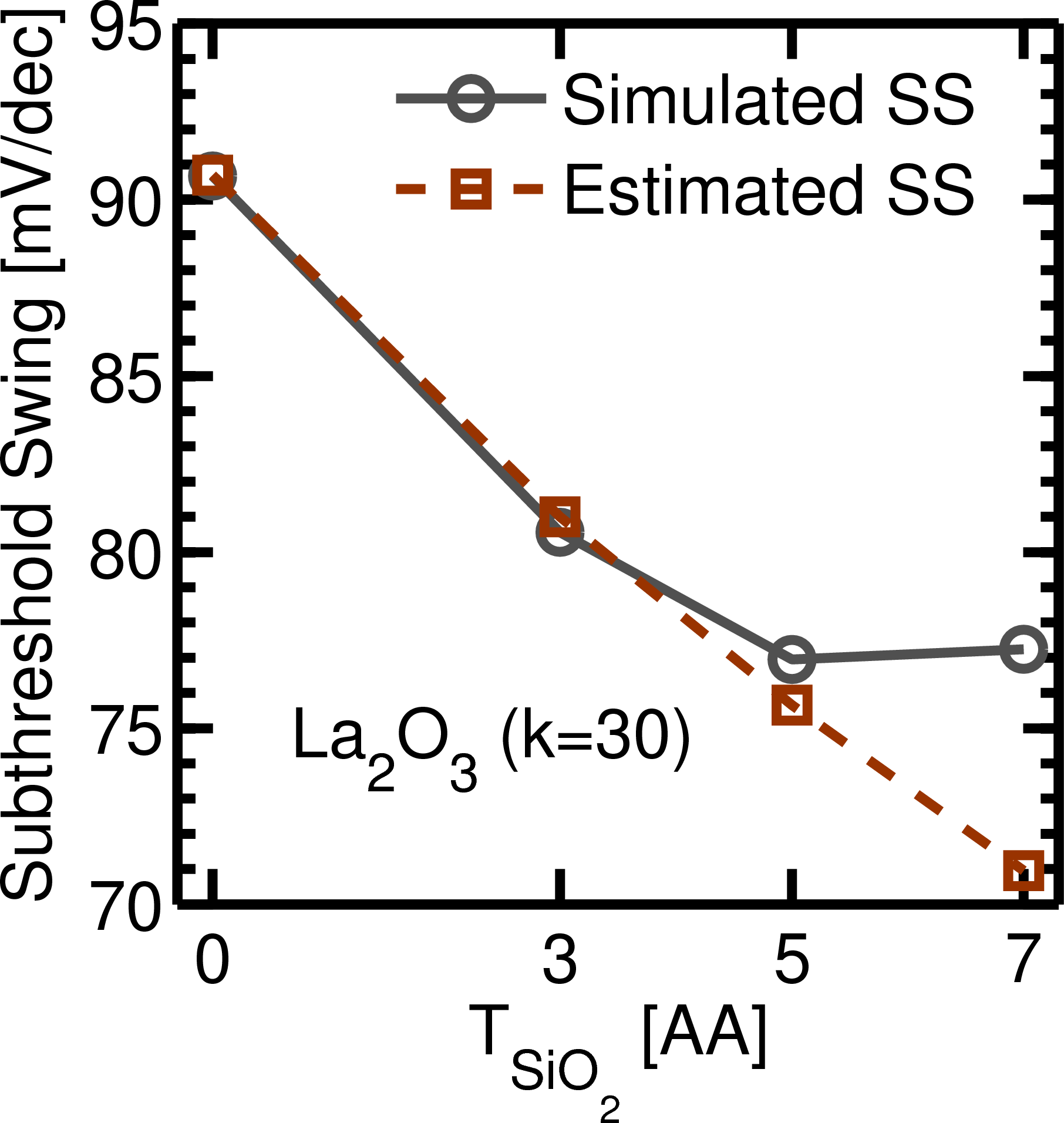}
		\put(20,60){(B)} 
		\end{overpic} \\           
  \end{tabular} 
\end{tabular}
\end{center}
\caption{\label{fig:ELECTROSTATICSALYSIS}
(A) Capacitance network between the gate and channel, source and drain is depicted. While fringing capacitance between the gate and the source/drain increases the total oxide capacitance degrades. The potential between gate and the channel and the source/drain rather than being confined only between the gate and the channel. This degrades the device performance for thicker gate stack. (B) Electrostatics degradation factor, $1-\frac{\alpha T_{OX}}{L_G}$, reduces by thinning the gate stack. This impact on SS for different $\rm La_2O3$ gate stack is depicted as an example. Analytically calculated SS is shown over simulated ones.}
\end{figure}

\textbf{CONCLUSION} - Scaling MOSFETs below 20 nm using thicker high-k oxides drastically degrades the device performance for a fixed EOT. Therefore, introducing higher-k oxide should be examined carefully for penalty in electrostatics. Using very thin oxides also causes gate leakage. An optimum combination of k value and oxide effective mass and thickness, or engineered gate stack, should be used to provide strong electrostatics and acceptable gate leakage. We provided a simple analytical model to estimate the gate leakage for each gate stack, as well as a model for the electrostatic degradation due to the fringing capacitance. Additionally, we showed there is an optimum gate stack thickness for any high-k material. Gate overlaps, source/drain doping and scattering can impact on the quantitative results. Therefore, for any specific device, it will be more accurate to do further investigation to find the optimum geometry and design for the best performance.  

\begin{acknowledgments}
The authors would like to thank M. Rush, Y. M. Tan, C.P. Chang, C. Cheung, J. Fonseca and S. Mehrotra. The use of nanoHUB.org computational resources operated by the Network for Computational Nanotechnology funded by the US National Science Foundation under Grant Nos. EEC-0228390, EEC-1227110, EEC-0228390, EEC-0634750, OCI-0438246, OCI-0832623 and OCI-0721680 is gratefully acknowledged. This work is also part of the "Accelerating Nano-scale Transistor Innovation with NEMO5 on Blue Waters" PRAC allocation support by the National Science Foundation (award number OCI-0832623).
\end{acknowledgments}

%\appendix

%\nocite{*}
\bibliography{aipsamp}% Produces the bibliography via BibTeX.

\providecommand{\noopsort}[1]{}\providecommand{\singleletter}[1]{#1}%
\begin{thebibliography}{20}%
\makeatletter
\providecommand \@ifxundefined [1]{%
 \@ifx{#1\undefined}
}%
\providecommand \@ifnum [1]{%
 \ifnum #1\expandafter \@firstoftwo
 \else \expandafter \@secondoftwo
 \fi
}%
\providecommand \@ifx [1]{%
 \ifx #1\expandafter \@firstoftwo
 \else \expandafter \@secondoftwo
 \fi
}%
\providecommand \natexlab [1]{#1}%
\providecommand \enquote  [1]{``#1''}%
\providecommand \bibnamefont  [1]{#1}%
\providecommand \bibfnamefont [1]{#1}%
\providecommand \citenamefont [1]{#1}%
\providecommand \href@noop [0]{\@secondoftwo}%
\providecommand \href [0]{\begingroup \@sanitize@url \@href}%
\providecommand \@href[1]{\@@startlink{#1}\@@href}%
\providecommand \@@href[1]{\endgroup#1\@@endlink}%
\providecommand \@sanitize@url [0]{\catcode `\\12\catcode `\$12\catcode
  `\&12\catcode `\#12\catcode `\^12\catcode `\_12\catcode `\%12\relax}%
\providecommand \@@startlink[1]{}%
\providecommand \@@endlink[0]{}%
\providecommand \url  [0]{\begingroup\@sanitize@url \@url }%
\providecommand \@url [1]{\endgroup\@href {#1}{\urlprefix }}%
\providecommand \urlprefix  [0]{URL }%
\providecommand \Eprint [0]{\href }%
\providecommand \doibase [0]{http://dx.doi.org/}%
\providecommand \selectlanguage [0]{\@gobble}%
\providecommand \bibinfo  [0]{\@secondoftwo}%
\providecommand \bibfield  [0]{\@secondoftwo}%
\providecommand \translation [1]{[#1]}%
\providecommand \BibitemOpen [0]{}%
\providecommand \bibitemStop [0]{}%
\providecommand \bibitemNoStop [0]{.\EOS\space}%
\providecommand \EOS [0]{\spacefactor3000\relax}%
\providecommand \BibitemShut  [1]{\csname bibitem#1\endcsname}%
\let\auto@bib@innerbib\@empty
%</preamble>
\bibitem [{ITR()}]{ITRS}%
  \BibitemOpen
  \href@noop {} {\enquote {\bibinfo {title} {Publications of international
  technology roadmap for semiconductors (itrs), 2013 editions.
  (http://www.itrs.net)},}\ }\BibitemShut {NoStop}%
\bibitem [{\citenamefont {Auth}\ \emph {et~al.}(2012)\citenamefont {Auth},
  \citenamefont {Allen}, \citenamefont {Blattner}, \citenamefont {Bergstrom},
  \citenamefont {Brazier}, \citenamefont {Bost}, \citenamefont {Buehler},
  \citenamefont {Chikarmane}, \citenamefont {Ghani}, \citenamefont {Glassman}
  \emph {et~al.}}]{IntelFinFET30nm}%
  \BibitemOpen
  \bibfield  {author} {\bibinfo {author} {\bibfnamefont {C.}~\bibnamefont
  {Auth}}, \bibinfo {author} {\bibfnamefont {C.}~\bibnamefont {Allen}},
  \bibinfo {author} {\bibfnamefont {A.}~\bibnamefont {Blattner}}, \bibinfo
  {author} {\bibfnamefont {D.}~\bibnamefont {Bergstrom}}, \bibinfo {author}
  {\bibfnamefont {M.}~\bibnamefont {Brazier}}, \bibinfo {author} {\bibfnamefont
  {M.}~\bibnamefont {Bost}}, \bibinfo {author} {\bibfnamefont {M.}~\bibnamefont
  {Buehler}}, \bibinfo {author} {\bibfnamefont {V.}~\bibnamefont {Chikarmane}},
  \bibinfo {author} {\bibfnamefont {T.}~\bibnamefont {Ghani}}, \bibinfo
  {author} {\bibfnamefont {T.}~\bibnamefont {Glassman}},  \emph {et~al.},\ }in\
  \href@noop {} {\emph {\bibinfo {booktitle} {VLSI Technology (VLSIT), 2012
  Symposium on}}}\ (\bibinfo {organization} {IEEE},\ \bibinfo {year} {2012})\
  pp.\ \bibinfo {pages} {131--132}\BibitemShut {NoStop}%
\bibitem [{\citenamefont {Salmani-Jelodar}\ \emph
  {et~al.}(2014{\natexlab{a}})\citenamefont {Salmani-Jelodar}, \citenamefont
  {Kim}, \citenamefont {Ng},\ and\ \citenamefont {Klimeck}}]{Salmani2014APL}%
  \BibitemOpen
  \bibfield  {author} {\bibinfo {author} {\bibfnamefont {M.}~\bibnamefont
  {Salmani-Jelodar}}, \bibinfo {author} {\bibfnamefont {S.}~\bibnamefont
  {Kim}}, \bibinfo {author} {\bibfnamefont {K.}~\bibnamefont {Ng}}, \ and\
  \bibinfo {author} {\bibfnamefont {G.}~\bibnamefont {Klimeck}},\ }\href@noop
  {} {\bibfield  {journal} {\bibinfo  {journal} {Applied Physics Letters}\
  }\textbf {\bibinfo {volume} {105}},\ \bibinfo {pages} {083508} (\bibinfo
  {year} {2014}{\natexlab{a}})}\BibitemShut {NoStop}%
\bibitem [{\citenamefont {Xie}\ \emph {et~al.}(2012)\citenamefont {Xie},
  \citenamefont {Xu},\ and\ \citenamefont {Taur}}]{xie2012review}%
  \BibitemOpen
  \bibfield  {author} {\bibinfo {author} {\bibfnamefont {Q.}~\bibnamefont
  {Xie}}, \bibinfo {author} {\bibfnamefont {J.}~\bibnamefont {Xu}}, \ and\
  \bibinfo {author} {\bibfnamefont {Y.}~\bibnamefont {Taur}},\ }\href@noop {}
  {\bibfield  {journal} {\bibinfo  {journal} {Electron Devices, IEEE
  Transactions on}\ }\textbf {\bibinfo {volume} {59}},\ \bibinfo {pages} {1569}
  (\bibinfo {year} {2012})}\BibitemShut {NoStop}%
\bibitem [{\citenamefont {Xie}\ \emph {et~al.}(2013)\citenamefont {Xie},
  \citenamefont {Lee}, \citenamefont {Xu}, \citenamefont {Wann}, \citenamefont
  {Sun},\ and\ \citenamefont {Taur}}]{xie2013comprehensive}%
  \BibitemOpen
  \bibfield  {author} {\bibinfo {author} {\bibfnamefont {Q.}~\bibnamefont
  {Xie}}, \bibinfo {author} {\bibfnamefont {C.-J.}\ \bibnamefont {Lee}},
  \bibinfo {author} {\bibfnamefont {J.}~\bibnamefont {Xu}}, \bibinfo {author}
  {\bibfnamefont {C.}~\bibnamefont {Wann}}, \bibinfo {author} {\bibfnamefont
  {J.-C.}\ \bibnamefont {Sun}}, \ and\ \bibinfo {author} {\bibfnamefont
  {Y.}~\bibnamefont {Taur}},\ }\href@noop {} {\bibfield  {journal} {\bibinfo
  {journal} {Electron Devices, IEEE Transactions on}\ }\textbf {\bibinfo
  {volume} {60}},\ \bibinfo {pages} {1814} (\bibinfo {year}
  {2013})}\BibitemShut {NoStop}%
\bibitem [{\citenamefont {Salmani-Jelodar}\ \emph
  {et~al.}(2014{\natexlab{b}})\citenamefont {Salmani-Jelodar}, \citenamefont
  {Kim}, \citenamefont {Ng},\ and\ \citenamefont {Klimeck}}]{Salmani2014SNW}%
  \BibitemOpen
  \bibfield  {author} {\bibinfo {author} {\bibfnamefont {M.}~\bibnamefont
  {Salmani-Jelodar}}, \bibinfo {author} {\bibfnamefont {S.}~\bibnamefont
  {Kim}}, \bibinfo {author} {\bibfnamefont {K.}~\bibnamefont {Ng}}, \ and\
  \bibinfo {author} {\bibfnamefont {G.}~\bibnamefont {Klimeck}},\ }in\
  \href@noop {} {\emph {\bibinfo {booktitle} {Silicon Nanoelectronics Workshop
  (SNW), 2014 IEEE}}}\ (\bibinfo {organization} {IEEE},\ \bibinfo {year}
  {2014})\ pp.\ \bibinfo {pages} {16.1--16.2}\BibitemShut {NoStop}%
\bibitem [{\citenamefont {Cheng}\ \emph {et~al.}(1999)\citenamefont {Cheng},
  \citenamefont {Cao}, \citenamefont {Rao}, \citenamefont {Inani},
  \citenamefont {Vande~Voorde}, \citenamefont {Greene}, \citenamefont {Stork},
  \citenamefont {Yu}, \citenamefont {Zeitzoff},\ and\ \citenamefont
  {Woo}}]{cheng1999impact}%
  \BibitemOpen
  \bibfield  {author} {\bibinfo {author} {\bibfnamefont {B.}~\bibnamefont
  {Cheng}}, \bibinfo {author} {\bibfnamefont {M.}~\bibnamefont {Cao}}, \bibinfo
  {author} {\bibfnamefont {R.}~\bibnamefont {Rao}}, \bibinfo {author}
  {\bibfnamefont {A.}~\bibnamefont {Inani}}, \bibinfo {author} {\bibfnamefont
  {P.}~\bibnamefont {Vande~Voorde}}, \bibinfo {author} {\bibfnamefont {W.~M.}\
  \bibnamefont {Greene}}, \bibinfo {author} {\bibfnamefont {J.~M.}\
  \bibnamefont {Stork}}, \bibinfo {author} {\bibfnamefont {Z.}~\bibnamefont
  {Yu}}, \bibinfo {author} {\bibfnamefont {P.~M.}\ \bibnamefont {Zeitzoff}}, \
  and\ \bibinfo {author} {\bibfnamefont {J.~C.}\ \bibnamefont {Woo}},\
  }\href@noop {} {\bibfield  {journal} {\bibinfo  {journal} {Electron Devices,
  IEEE Transactions on}\ }\textbf {\bibinfo {volume} {46}},\ \bibinfo {pages}
  {1537} (\bibinfo {year} {1999})}\BibitemShut {NoStop}%
\bibitem [{\citenamefont {Frank}\ \emph {et~al.}(1998)\citenamefont {Frank},
  \citenamefont {Taur},\ and\ \citenamefont {Wong}}]{frank1998generalized}%
  \BibitemOpen
  \bibfield  {author} {\bibinfo {author} {\bibfnamefont {D.~J.}\ \bibnamefont
  {Frank}}, \bibinfo {author} {\bibfnamefont {Y.}~\bibnamefont {Taur}}, \ and\
  \bibinfo {author} {\bibfnamefont {H.-S.}\ \bibnamefont {Wong}},\ }\href@noop
  {} {\bibfield  {journal} {\bibinfo  {journal} {Electron Device Letters,
  IEEE}\ }\textbf {\bibinfo {volume} {19}},\ \bibinfo {pages} {385} (\bibinfo
  {year} {1998})}\BibitemShut {NoStop}%
\bibitem [{\citenamefont {Manoj}\ and\ \citenamefont
  {Ramgopal~Rao}(2007)}]{manoj2007impact}%
  \BibitemOpen
  \bibfield  {author} {\bibinfo {author} {\bibfnamefont {C.}~\bibnamefont
  {Manoj}}\ and\ \bibinfo {author} {\bibfnamefont {V.}~\bibnamefont
  {Ramgopal~Rao}},\ }\href@noop {} {\bibfield  {journal} {\bibinfo  {journal}
  {Electron Device Letters, IEEE}\ }\textbf {\bibinfo {volume} {28}},\ \bibinfo
  {pages} {295} (\bibinfo {year} {2007})}\BibitemShut {NoStop}%
\bibitem [{\citenamefont {Cho}\ \emph {et~al.}(2011)\citenamefont {Cho},
  \citenamefont {Seo}, \citenamefont {Jeong}, \citenamefont {Kim},
  \citenamefont {Lim}, \citenamefont {Jang}, \citenamefont {Hong},
  \citenamefont {Suk}, \citenamefont {Li}, \citenamefont {Ryou} \emph
  {et~al.}}]{Samsung2011Bulk}%
  \BibitemOpen
  \bibfield  {author} {\bibinfo {author} {\bibfnamefont {H.-J.}\ \bibnamefont
  {Cho}}, \bibinfo {author} {\bibfnamefont {K.-I.}\ \bibnamefont {Seo}},
  \bibinfo {author} {\bibfnamefont {W.}~\bibnamefont {Jeong}}, \bibinfo
  {author} {\bibfnamefont {Y.-H.}\ \bibnamefont {Kim}}, \bibinfo {author}
  {\bibfnamefont {Y.}~\bibnamefont {Lim}}, \bibinfo {author} {\bibfnamefont
  {W.}~\bibnamefont {Jang}}, \bibinfo {author} {\bibfnamefont {J.}~\bibnamefont
  {Hong}}, \bibinfo {author} {\bibfnamefont {S.}~\bibnamefont {Suk}}, \bibinfo
  {author} {\bibfnamefont {M.}~\bibnamefont {Li}}, \bibinfo {author}
  {\bibfnamefont {C.}~\bibnamefont {Ryou}},  \emph {et~al.},\ }in\ \href@noop
  {} {\emph {\bibinfo {booktitle} {Electron Devices Meeting (IEDM), 2011 IEEE
  International}}}\ (\bibinfo {organization} {IEEE},\ \bibinfo {year} {2011})\
  pp.\ \bibinfo {pages} {15.1.1 -- 15.1.4}\BibitemShut {NoStop}%
\bibitem [{\citenamefont {Luisier}\ \emph {et~al.}(2007)\citenamefont
  {Luisier}, \citenamefont {Schenk},\ and\ \citenamefont
  {Fichtner}}]{luisier2007three}%
  \BibitemOpen
  \bibfield  {author} {\bibinfo {author} {\bibfnamefont {M.}~\bibnamefont
  {Luisier}}, \bibinfo {author} {\bibfnamefont {A.}~\bibnamefont {Schenk}}, \
  and\ \bibinfo {author} {\bibfnamefont {W.}~\bibnamefont {Fichtner}},\ }in\
  \href@noop {} {\emph {\bibinfo {booktitle} {Electron Devices Meeting, 2007.
  IEDM 2007. IEEE International}}}\ (\bibinfo {organization} {IEEE},\ \bibinfo
  {year} {2007})\ pp.\ \bibinfo {pages} {733--736}\BibitemShut {NoStop}%
\bibitem [{\citenamefont {Luisier}\ and\ \citenamefont
  {Schenk}(2008)}]{luisier2008two}%
  \BibitemOpen
  \bibfield  {author} {\bibinfo {author} {\bibfnamefont {M.}~\bibnamefont
  {Luisier}}\ and\ \bibinfo {author} {\bibfnamefont {A.}~\bibnamefont
  {Schenk}},\ }\href@noop {} {\bibfield  {journal} {\bibinfo  {journal}
  {Electron Devices, IEEE Transactions on}\ }\textbf {\bibinfo {volume} {55}},\
  \bibinfo {pages} {1494} (\bibinfo {year} {2008})}\BibitemShut {NoStop}%
\bibitem [{\citenamefont {Steiger}\ \emph {et~al.}(2011)\citenamefont
  {Steiger}, \citenamefont {Povolotskyi}, \citenamefont {Park}, \citenamefont
  {Kubis},\ and\ \citenamefont {Klimeck}}]{steiger2011nemo5}%
  \BibitemOpen
  \bibfield  {author} {\bibinfo {author} {\bibfnamefont {S.}~\bibnamefont
  {Steiger}}, \bibinfo {author} {\bibfnamefont {M.}~\bibnamefont
  {Povolotskyi}}, \bibinfo {author} {\bibfnamefont {H.-H.}\ \bibnamefont
  {Park}}, \bibinfo {author} {\bibfnamefont {T.}~\bibnamefont {Kubis}}, \ and\
  \bibinfo {author} {\bibfnamefont {G.}~\bibnamefont {Klimeck}},\ }\href@noop
  {} {\bibfield  {journal} {\bibinfo  {journal} {Nanotechnology, IEEE
  Transactions on}\ }\textbf {\bibinfo {volume} {10}},\ \bibinfo {pages} {1464}
  (\bibinfo {year} {2011})}\BibitemShut {NoStop}%
\bibitem [{\citenamefont {Fonseca}\ \emph {et~al.}(2013)\citenamefont
  {Fonseca}, \citenamefont {Kubis}, \citenamefont {Povolotskyi}, \citenamefont
  {Novakovic}, \citenamefont {Ajoy}, \citenamefont {Hegde}, \citenamefont
  {Ilatikhameneh}, \citenamefont {Jiang}, \citenamefont {Sengupta},
  \citenamefont {Tan} \emph {et~al.}}]{fonseca2013efficient}%
  \BibitemOpen
  \bibfield  {author} {\bibinfo {author} {\bibfnamefont {J.~E.}\ \bibnamefont
  {Fonseca}}, \bibinfo {author} {\bibfnamefont {T.}~\bibnamefont {Kubis}},
  \bibinfo {author} {\bibfnamefont {M.}~\bibnamefont {Povolotskyi}}, \bibinfo
  {author} {\bibfnamefont {B.}~\bibnamefont {Novakovic}}, \bibinfo {author}
  {\bibfnamefont {A.}~\bibnamefont {Ajoy}}, \bibinfo {author} {\bibfnamefont
  {G.}~\bibnamefont {Hegde}}, \bibinfo {author} {\bibfnamefont
  {H.}~\bibnamefont {Ilatikhameneh}}, \bibinfo {author} {\bibfnamefont
  {Z.}~\bibnamefont {Jiang}}, \bibinfo {author} {\bibfnamefont
  {P.}~\bibnamefont {Sengupta}}, \bibinfo {author} {\bibfnamefont
  {Y.}~\bibnamefont {Tan}},  \emph {et~al.},\ }\href@noop {} {\bibfield
  {journal} {\bibinfo  {journal} {Journal of Computational Electronics}\
  }\textbf {\bibinfo {volume} {12}},\ \bibinfo {pages} {592} (\bibinfo {year}
  {2013})}\BibitemShut {NoStop}%
\bibitem [{\citenamefont {Luisier}\ \emph {et~al.}(2006)\citenamefont
  {Luisier}, \citenamefont {Schenk}, \citenamefont {Fichtner},\ and\
  \citenamefont {Klimeck}}]{luisier2006atomistic}%
  \BibitemOpen
  \bibfield  {author} {\bibinfo {author} {\bibfnamefont {M.}~\bibnamefont
  {Luisier}}, \bibinfo {author} {\bibfnamefont {A.}~\bibnamefont {Schenk}},
  \bibinfo {author} {\bibfnamefont {W.}~\bibnamefont {Fichtner}}, \ and\
  \bibinfo {author} {\bibfnamefont {G.}~\bibnamefont {Klimeck}},\ }\href@noop
  {} {\bibfield  {journal} {\bibinfo  {journal} {Physical Review B}\ }\textbf
  {\bibinfo {volume} {74}},\ \bibinfo {pages} {205323} (\bibinfo {year}
  {2006})}\BibitemShut {NoStop}%
\bibitem [{\citenamefont {Salmani-Jelodar}\ \emph
  {et~al.}(2014{\natexlab{c}})\citenamefont {Salmani-Jelodar}, \citenamefont
  {Mehrotra}, \citenamefont {Ilatikhameneh},\ and\ \citenamefont
  {Klimeck}}]{salmaniTunneling}%
  \BibitemOpen
  \bibfield  {author} {\bibinfo {author} {\bibfnamefont {M.}~\bibnamefont
  {Salmani-Jelodar}}, \bibinfo {author} {\bibfnamefont {S.}~\bibnamefont
  {Mehrotra}}, \bibinfo {author} {\bibfnamefont {H.}~\bibnamefont
  {Ilatikhameneh}}, \ and\ \bibinfo {author} {\bibfnamefont {G.}~\bibnamefont
  {Klimeck}},\ }\href@noop {} {\bibfield  {journal} {\bibinfo  {journal}
  {Nanotechnology, IEEE Transactions on}\ }\textbf {\bibinfo {volume} {accepted
  for publication}} (\bibinfo {year} {2014}{\natexlab{c}})}\BibitemShut
  {NoStop}%
\bibitem [{\citenamefont {Yeo}\ \emph {et~al.}(2002)\citenamefont {Yeo},
  \citenamefont {King},\ and\ \citenamefont {Hu}}]{yeo2002direct}%
  \BibitemOpen
  \bibfield  {author} {\bibinfo {author} {\bibfnamefont {Y.-C.}\ \bibnamefont
  {Yeo}}, \bibinfo {author} {\bibfnamefont {T.-J.}\ \bibnamefont {King}}, \
  and\ \bibinfo {author} {\bibfnamefont {C.}~\bibnamefont {Hu}},\ }\href@noop
  {} {\bibfield  {journal} {\bibinfo  {journal} {Applied Physics Letters}\
  }\textbf {\bibinfo {volume} {81}},\ \bibinfo {pages} {2091} (\bibinfo {year}
  {2002})}\BibitemShut {NoStop}%
\bibitem [{\citenamefont {Robertson}(2000)}]{robertson2000band}%
  \BibitemOpen
  \bibfield  {author} {\bibinfo {author} {\bibfnamefont {J.}~\bibnamefont
  {Robertson}},\ }\href@noop {} {\bibfield  {journal} {\bibinfo  {journal}
  {Journal of Vacuum Science \& Technology B}\ }\textbf {\bibinfo {volume}
  {18}},\ \bibinfo {pages} {1785} (\bibinfo {year} {2000})}\BibitemShut
  {NoStop}%
\bibitem [{\citenamefont {Datta}(2005)}]{datta2005quantum}%
  \BibitemOpen
  \bibfield  {author} {\bibinfo {author} {\bibfnamefont {S.}~\bibnamefont
  {Datta}},\ }\href@noop {} {\emph {\bibinfo {title} {Quantum transport: atom
  to transistor}}}\ (\bibinfo  {publisher} {Cambridge University Press},\
  \bibinfo {year} {2005})\BibitemShut {NoStop}%
\bibitem [{\citenamefont {Sze}\ and\ \citenamefont
  {Ng}(2006)}]{sze2006physics}%
  \BibitemOpen
  \bibfield  {author} {\bibinfo {author} {\bibfnamefont {S.~M.}\ \bibnamefont
  {Sze}}\ and\ \bibinfo {author} {\bibfnamefont {K.~K.}\ \bibnamefont {Ng}},\
  }\href@noop {} {\emph {\bibinfo {title} {Physics of semiconductor devices}}}\
  (\bibinfo  {publisher} {John Wiley \& Sons},\ \bibinfo {year}
  {2006})\BibitemShut {NoStop}%
\end{thebibliography}%

\end{document}